\documentstyle[12pt]{article}
\begin{document}
\pagestyle{plain}
\setcounter{page}{1}
\baselineskip=0.3in
\begin{titlepage}
\begin{flushright}
NUHEP-TH-97-8\\
hep-ph/9706412
\end{flushright}
\vspace{.5cm}

\begin{center}
{\Large Supersymmetric QCD Corrections to Single Top Quark \\
                Production at the Fermilab Tevatron }

\vspace{.2in}

     Chong Sheng Li $^a$, Robert J. Oakes $^b$, Jin Min Yang$^{b,}$
     \footnote{ On leave from Department of Physics, Henan Normal University, 
      China},
     and Hong-Yi Zhou $^c$
\vspace{.2in}

$^a$ Department of Physics, Peking University,\\
     Beijing 100871, China\\
$^b$ Department of Physics and Astronomy, Northwestern University,\\
     Evanston, Illinois 60208, USA\\
$^c$ Institute of Modern Physics, Tsinghua University,\\
     Beijing 100034, China
\end{center}
\vspace{.4in}

\begin{footnotesize}
\begin{center}\begin{minipage}{5in}
\baselineskip=0.25in
\begin{center} ABSTRACT\end{center}
                    
  We evaluate  the supersymmetric QCD corrections to single top quark 
production via $q \bar q'\rightarrow t \bar b$
at the Fermilab Tevatron in the minimal supersymmetric model.
We find that within the  allowed range of squark and gluino masses
the supersymmetric QCD corrections can enhance the cross section
by a few percent. 
The combined effects of SUSY QCD, SUSY EW, and the Yukawa couplings
can exceed 10\%  for the smallest allowed $\tan\beta~ (\simeq 0.25)$ 
but are only a few percent for $\tan\beta >1$. 
\end{minipage}\end{center}
\end{footnotesize}
\vfill


\end{titlepage}
\eject
\baselineskip=0.25in
\begin{center} {\Large 1. Introduction }\end{center}

Even with fewer events expected, single top quark production at the Tevatron 
is also
important because it involves the electroweak interaction and, therefore,
can probe the electroweak sector of the theory, in contrast to the dominant
QCD pair
production mechanism, and provide a consistency check on the measured
parameters of the top quark in the QCD pair production process. At the Tevatron
single top quarks are produced primarily via the $W$-gluon fusion process[1]
and the quark annihilation process, 
$q \bar q'\rightarrow W^* \rightarrow t\bar b$ ($W^*$ process)[2], 
which can reliably be predicted in the Standard Model (SM), 
and the theoretical uncertainty in the cross section 
is only about a few percent due to QCD corections [3].
As shown in Ref.[4],
 a high-luminosity Tevatron would allow a measurement of this cross section 
with a statistical uncertainty of about 6\%. At this level of experimental 
accuracy a calculation of the radiative corrections is necessary 
to compare with the SM and to look for physics beyond the SM.

In Ref.[4] the QCD and Yukawa corrections to the $W^*$ process
have been calculated in the SM. 
In the Minimal Supersymmetric Model(MSSM)[5],
the Yukawa corrections from the Higgs sector and the electroweak 
corrections from chargino and neutralino couplings
have also been evaluated [6,7].
Very recently the effects of R-parity-violating couplings in this process
were investigated[8]. In addition to these effects the SUSY QCD corrections
may also be significant and need to be included, also.
In this paper we evaluate the SUSY QCD corrections to single top production 
from the $W^*$ process at the Fermilab Tevatron in the MSSM.
In Sec. II we present the analytic results
in terms of the well-known standard notation of one-loop Feynman integrals.
In Sec. III we give some numerical examples and discuss the implications
of our results.
\vspace{1cm}

\begin{center} {\Large 2. Calculations }\end{center}
\vspace{.3cm}

  The tree-level Feynman diagram for single top quark production via
the $W^*$ process, $q \bar q'\rightarrow t \bar b$, is shown in Fig.1(a).
The SUSY QCD contributions to the amplitude are contained in the corrections
to the  $Wq\bar q'$ and $Wt\bar b$ vertices where the $W$ boson is off-shell. 
The  relavent Feynman diagrams are shown in Figs.1(b-g). 
Note that the SUSY QCD contributions to the $Wt\bar b$ vertex  
with all the particles on-shell were studied in Refs.[9,10].
In our calculations we used dimensional regularization to 
control all the ultraviolet divergences in the virtual loop corrections 
and we adopted the on-mass-shell renormalization scheme[11].
Including the SUSY QCD corrections, the renormalized amplitude for
$q\bar q'\rightarrow t\bar b$ can be written as
\begin{equation}
M_{ren}=M_0 +\delta M
\end{equation}
where $M_0$ is  the tree-level matrix element and $\delta M$ 
represents the SUSY QCD corrections.
 $M_0$ is given by
\begin{equation}
 M_0= i \frac{g^2}{2} \frac{1}{\hat s-m_W^2}
  \bar v(p_2) \gamma_{\mu} P_L u(p_1)  \bar u(p_3) \gamma^{\mu} P_L v(p_4),
\end{equation} 
where $p_1$ and $p_2$ denote the momentum of the incoming quarks 
$q$ and $\bar q'$, while $p_3$ and $p_4$ refer to the outgoing $t$ and
$\bar b$ quarks, and $\hat{s}$ is the center-of-mass energy of the
subprocess. 
$\delta M$  is given by
\begin{equation}
\delta M=\delta M_{Wt\bar b}+\delta M_{Wq\bar q'}
\end{equation}
where $\delta M_{Wt\bar b}$ and $\delta M_{Wq\bar q'}$
represent the corrections to the $Wt\bar b$ and $Wq\bar q'$ vertices, 
respectively.
Calculating the vertex and self-energy diagrams we find
\begin{eqnarray}
\delta M_{Wt\bar b}&= &i \frac{g^2}{2} \frac{1}{\hat s-m_W^2}
  \bar v(p_2) \gamma_{\mu} P_L u(p_1)~
  \bar u(p_3) \left [\gamma^{\mu} P_L (\frac{1}{2}\delta Z_t^L+
\frac{1}{2}\delta Z_b^L+f^L_1)\right.
        \nonumber\\
& &  \left.+\gamma^{\mu} P_R f^R_1+p^{\mu}_3 P_L f^L_2+ p^{\mu}_4 P_L f^L_3
        +p^{\mu}_3 P_R f^R_2+ p^{\mu}_4 P_R f^R_3\right ] v(p_4),
\end{eqnarray}
and 
\begin{eqnarray}
\delta M_{Wq\bar q'}&= &i \frac{g^2}{2} \frac{1}{\hat s-m_W^2}
  \bar u(p_3) \gamma_{\mu} P_L v(p_4)~
  \bar v(p_2) \left [\gamma^{\mu} P_L (\frac{1}{2}\delta Z_q^L+
\frac{1}{2}\delta Z_{q'}^L+f'^L_1)\right.
        \nonumber\\
& &  \left.+\gamma^{\mu} P_R f'^R_1+p^{\mu}_1 P_L f'^L_2+ p^{\mu}_2 P_L f'^L_3
        +p^{\mu}_1 P_R f'^R_2+ p^{\mu}_2 P_R f'^R_3\right ] u(p_1).
\end{eqnarray}
Here  the renormalization constants $\delta Z^L_q (q=t,b,q,q')$ 
and the form factors $f^L_{1,2,3}$ are 
\begin{eqnarray}
\delta Z^L_q &=&\frac{\alpha_s C_F}{4\pi}\left [
        (a_{\tilde q_i}-b_{\tilde q_i})^2 (-\frac{\Delta}{2}
        +F_1^{(q\tilde g \tilde q_i)}
        +2m_q^2(a_{\tilde q_i}^2+b_{\tilde q_i}^2)G_1^{(q\tilde g \tilde q_i)}
        \right. \nonumber \\
& &\left. +2m_qm_{\tilde g}(a_{\tilde q_i}^2-b_{\tilde q_i}^2)
        G_0^{(q\tilde g \tilde q_i)}\right ],\\
f^L_1&=&\frac{\alpha_s C_F}{2\pi}\alpha_{\tilde t_i\tilde b_j}
        \lambda_{\tilde t_i\tilde b_j} c_{24},\\
f^L_2&=&-\frac{\alpha_s C_F}{4\pi}\alpha_{\tilde t_i\tilde b_j}
        \left [ m_{\tilde g}\eta'_{\tilde t_i\tilde b_j} (c_0+2c_{11}-2c_{12})
        \right.\nonumber\\
         & & \left.+m_t \lambda_{\tilde t_i\tilde b_j}(c_{12}-c_{11}-2c_{21}
                -2c_{22}+4c_{23})\right ],
\end{eqnarray}
and
\begin{eqnarray}
f^L_3&=&-\frac{\alpha_s C_F}{4\pi}\alpha_{\tilde t_i\tilde b_j}
        \left [-m_{\tilde g}\eta'_{\tilde t_i\tilde b_j} (c_0+2c_{12})
               +m_t \lambda_{\tilde t_i\tilde b_j}(c_{11}-c_{12}-2c_{22}
                +2c_{23})\right ],
\end{eqnarray}
where the color factor 
$C_F=4/3$ and sums over $i,j=1,2$ are implied.
The functions $c_{ij}(-p_3,p_3+p_4,m_{\tilde g},m_{\tilde t_i},m_{\tilde b_j})$
in $f^L_{n}$, are the usual Feynman integrals [14].
The functions $F^{(ijk)}_{0,1}$ and $G^{(ijk)}_{0,1}$ are defined to be
\begin{eqnarray}
F^{(ijk)}_n&=&\int^1_0 dy y^n\log \left [\frac{m_i^2y(y-1)+m^2_j(1-y)
+m^2_k y}{\mu ^2}\right ],
\end{eqnarray}
and
\begin{eqnarray}
G^{(ijk)}_n&=&-\int^1_0 dy \frac{y^{n+1}(1-y)}{m_i^2y(y-1)+
m^2_j(1-y)+m^2_ky},
\end{eqnarray}
The other form factors can be obtained through the following substitutions:
\begin{eqnarray}
f^R_{1,2,3}&=&f^L_{1,2,3}\left \vert_{
\lambda_{\tilde t_i\tilde b_j}\rightarrow \eta_{\tilde t_i\tilde b_j},~
\eta'_{\tilde t_i\tilde b_j}\rightarrow \lambda'_{\tilde t_i\tilde b_j}},
 \right.\\
f'^{L,R}_n&=&f^{L,R}_n\left \vert_{
        \tilde t_i\tilde b_j\rightarrow \tilde q'_j\tilde q_i,~
        m_t\rightarrow 0,~p_3\rightarrow p_1,
        ~m_{\tilde t_i}\rightarrow m_{\tilde q_i},
        ~m_{\tilde b_j}\rightarrow m_{\tilde q'_j}},\right.
\end{eqnarray}
The constants $a_{\tilde q_i}, b_{\tilde q_i},
\alpha_{\tilde q_i\tilde q'_j},
\eta_{\tilde q_i\tilde q'_j},\eta'_{\tilde q_i\tilde q'_j},
\lambda_{\tilde q_i\tilde q'_j}$ and $\lambda'_{\tilde q_i\tilde q'_j}$
appearing above are defined by
\begin{eqnarray}
a_{\tilde q_1}&=&-b_{\tilde q_2}=\frac{1}{\sqrt 2}(\cos\theta_{\tilde q}
                -\sin\theta_{\tilde q}),\\
a_{\tilde q_2}&=&b_{\tilde q_1}=-\frac{1}{\sqrt 2}(\cos\theta_{\tilde q}
                +\sin\theta_{\tilde q}),\\
\alpha_{\tilde q_1\tilde q'_1}&=&\cos\theta_{\tilde q}\cos\theta_{\tilde q'},\\
\alpha_{\tilde q_2\tilde q'_2}&=&\sin\theta_{\tilde q}\sin\theta_{\tilde q'},\\
\alpha_{\tilde q_1\tilde q'_2}&=&-\cos\theta_{\tilde q}\sin\theta_{\tilde q'},\\
\alpha_{\tilde q_2\tilde q'_1}&=&-\sin\theta_{\tilde q}\cos\theta_{\tilde q'},\\
\eta_{\tilde q_i\tilde q'_j}&=&(a_{\tilde q_i}+b_{\tilde q_i})
                (a_{\tilde q'_j}+b_{\tilde q'_j}),\\
\eta'_{\tilde q_i\tilde q'_j}&=&(a_{\tilde q_i}+b_{\tilde q_i})
                (a_{\tilde q'_j}-b_{\tilde q'_j}),\\
\lambda_{\tilde q_i\tilde q'_j}&=&(a_{\tilde q_i}-b_{\tilde q_i})
                (a_{\tilde q'_j}-b_{\tilde q'_j}),
\end{eqnarray}
and 
\begin{eqnarray}
\lambda'_{\tilde q_i\tilde q'_j}&=&(a_{\tilde q_i}-b_{\tilde q_i})
                (a_{\tilde q'_j}+b_{\tilde q'_j}),~~~~~~~~~~
\end{eqnarray}
where $\theta_{\tilde q}$ is the mixing angle of the left- and right-handed
squarks $\tilde q_L$ and $\tilde q_R$.  

  The renormalized differential cross section for the subprocess is 
\begin{equation}
\frac{d\hat{\sigma}}{d\cos\theta}=\frac{\hat{s}-m_t^2}{32\pi\hat s^2}
\overline{\sum} \vert M_{ren}\vert^2,
\end{equation}
where $\theta$ is the angle between the top quark and incident initial
quark.
Integrating this differential cross section over $\cos\theta$ 
one finds the cross section for subprocess is of the form
\begin{equation}
\hat{\sigma}=\hat{\sigma}_0+\Delta \hat{\sigma}
\end{equation}
where the tree-level cross section $\hat{\sigma}_0$ is given by
\begin{equation}
\hat{\sigma}_0=\frac{g^4}{128\pi}\frac{\hat{s}-m^2_t}{\hat{s}^2(\hat{s}-
m^2_W)^2}[\frac{2}{3}(\hat{s}-m_t^2)^2
 +(\hat{s}-m^2_t)(m^2_t+m^2_b) + 2m^2_tm^2_b],
\end{equation}
and $\Delta \hat{\sigma}$ contains the SUSY QCD corrections to the subprocess
coming from the vertex corrections to the quark annihilation diagrams.

The total hadronic cross section for the production of a single top quark via 
$q\bar q'$ annihilation, the $W^*$ process, can be written in the form
\begin{equation}
\sigma (s)=\sum_{i,j}\int dx_1 dx_2 \hat\sigma_{ij}(x_1x_2s, m_t^2,
\mu^2)[f^A_i(x_1,\mu)f^B_j(x_2, \mu)+(A\leftrightarrow B)],
\end{equation}
where
\begin{eqnarray}
s&=&(P_1+P_2)^2,\\
\hat{s}&=&x_1x_2s,\\
p_1&=&x_1P_1,
\end{eqnarray}
and 
\begin{equation}
p_2=x_2P_2.
\end{equation}
Here $A$ and $B$ denote the incident hadrons and $P_1$ and $P_2$ are their
four-momenta, while $i, j=1,2$ denote the initial partons and $x_1$ and 
$x_2$ are 
their respective longitudinal momentum fractions. The functions $f^A_i$ and $f^B_j$ are 
the usual parton distributions[11,12]. 
Finally, introducing the convenient variable 
$\tau =x_1x_2$ and changing independent variables, the total cross section
becomes
\begin{equation}
\sigma(s)=\sum_{i,j}\int^1_{\tau_0}\frac{d\tau}{\tau}(\frac{1}{s}
\frac{dL_{ij}}{d\tau})(\hat s \hat \sigma_{ij})
\end{equation}
where $\tau_0=(m_t+m_b)^2/s$. The quantity $dL_{ij}/d\tau$ is the parton
luminosity which is defined to be
\begin{equation}
\frac{dL_{ij}}{d\tau}=\int^1_{\tau} \frac{dx_1}{x_1}[f^A_i(x_1,\mu)
f^B_j(\tau/x_1,\mu)+(A\leftrightarrow B)]
\end{equation}
\vspace{.4cm}

\begin{center} {\Large 3. Numerical results and conclusion }\end{center}

In the following we present numerical results for
the SUSY QCD corrections $\Delta \sigma$ to the total cross section for single
 top quark production via the $W^*$ subprocess $q \bar q'\rightarrow t \bar b$ 
at the Fermilab Tevatron assuming $\sqrt s=2$ TeV.
In our numerical calculations we used the MRSA$^{\prime}$ parton distribution
functions[13]. For the relevant parameters we choose $m_W=80.3 \rm{GeV},
m_t=176 \rm{GeV}, m_b=4.9 \rm{GeV}$ and $\alpha_{ew}=1/128.8$.
Also the cuts $\vert \eta \vert <2.5$ and $p_T>20$ GeV were assumed.

In our anlytical results several different squarks are involved:
$\tilde t_i$, $\tilde b_i$, $\tilde q_i$ and $\tilde q'_i$.   
The mixing between left- and right-handed squarks is negligible 
except for the stops.
However, in our calculation, for simplicity, we considered the special 
case: no mixing between left- and right-handed stops and degenerate
squark masses. Thus the SUSY parameters involved
in our calculation are then reduced to only two: the gluino mass
and the squark mass. This special case, while not including all
possible special values of the SUSY parameters, suffices to 
illustrate our main conclusions below.

Figure 2 shows the SUSY QCD correction $\Delta\sigma/\sigma_0$  
as a function of  gluino mass assuming a squark mass of 100GeV.
The correction is positive except for gluino masses  in the range of 
76 GeV$<m_{\tilde g}<$100 GeV.
Note that there is  a  peak at $m_{\tilde g}=76$ GeV due to the fact 
that for $m_t=176$ GeV the threshold for open top decay into a
gluino and a stop is crossed in this region. Only in this very narrow
range does the magnitude of the correction exceed
5\%. For the gluino mass range allowed by Tevatron experiments, 
, $m_{\tilde g}>170$ GeV [15],  the correction is only a couple of
percent.

Figure 3 presents the SUSY QCD correction $\Delta\sigma/\sigma_0$  as a 
function of the stop mass assuming $m_{\tilde g}=200$ GeV.
This correction is always positive. With increasing squark mass the
magnitude of the correction decreases, illustrating the decoupling effect. 
The corrections reach a couple of percent for a squark mass around 100 GeV
and are below one percent for squark masses above about 200 GeV.

As shown in Fig.4 of Ref.[7], for the minimum allowed 
$\tan\beta~ (\simeq 0.25)$,
the Yukawa corrections can enhance the cross section
by 10\%,  while the electroweak corrections can decrease
the cross section by more than 20\%.  
However, for $\tan\beta>1$ the Yukawa corrections are below 1\%
and electroweak corrections are about -4\%. As shown in Fig.2, the
SUSY QCD corrections are below 2\% for gluino masses larger than 
about 100 GeV. Therefore, for $\tan\beta>1$ and gluino masses larger than 
100 GeV, the combined effects of SUSY QCD, 
SUSY EW, and the Yukawa corrections only amount at most to a few percent,
 which will be difficult to detect at the Tevatron.
Note, however, that the R-parity-violating couplings in the MSSM can 
give rise to 
observable effects at the future upgraded Tevatron, as shown in Ref.[8].
In the R-parity-violating MSSM the total effect of SUSY
is the sum of all these contributions. 
To establish precise constraints on the R-parity-violating couplings
one will have to take into account all the contributions, which 
will presumably be necessary for the upgraded Tevatron data.
\vspace{.5cm}

This work was supported in part by the U.S. Department of Energy, Division
of High Energy Physics, under Grant No. DE-FG02-91-ER4086. 
\eject

{\LARGE References}
\vspace{0.2cm}
\begin{itemize}
\begin{description}
\item[{\rm[1]}] S. Dawson, Nucl. Phys. B249, 42 (1985)
                S. Willenbrock and D. Dicus, Phys. Rev. D34, 155 (1986);
                S. Dawson and S. Willenbrock, Nucl. Phys. B284, 449 (1987);
                C. P. Yuan, Phys. Rev. D41, 42 (1990);
            F. Anselmo, B. van Eijk and G. Bordes, Phys. Rev. D45, 2312 (1992);
                R. K. Ellis and S. Parke, Phys. Rev. D46, 3785 (1992);
                D. Carlson and C. P. Yuan, Phys. Lett. B306, 386 (1993);
                G. Bordes and B. van Eijk, Nucl. Phys. B435, 23 (1995);
                A. Heinson, A. Belyaev and E. Boos, hep-ph/9509274. 
\item[{\rm [2]}]  S. Cortese and R. Petronzio, Phys. Lett. B306, 386 (1993). 
\item[{\rm [3]}]  T. Stelzer and S. Willenbrock, Phys. Lett. B357, 125 (1995). 
\item[{\rm [4]}]  M. Smith and S. Willenbrock, hep-ph/9604223. 
\item[{\rm [5]}] H. E. Haber and C. L. Kane, Phys. Rep. 117, 75 (1985);
                 J. F. Gunion and H. E. Haber, Nucl. Phys. B272, 1 (1986). 
\item[{\rm [6]}] C. S. Li, R. J. Oakes and J. M. Yang, Phys. Rev. D55, 1672 (1997). 
\item[{\rm [7]}] C. S. Li, R. J. Oakes and J. M. Yang, Phys. Rev. D55, 5780 (1997). 
\item[{\rm [8]}] A. Datta, J. M. Yang, B. -L. Young and X. Zhang, hep-ph/9704257,
                 to appear in Phys. Rev. D. 
\item[{\rm [9]}]   C. S. Li, J. M. Yang and B. Q. Hu, Phys. Rev. D48, 5425 (1993). 
\item[{\rm [10]}]  A. Dabelstein, W. Hollik, C. Junger,
                   R. A. Jimenez and J. Sola, Nucl. Phys. B454, 75 (1995).  
\item[{\rm [11]}] A. Sirlin, Phys. Rev. D22, 971 (1980);
                W. J. Marciano and A. Sirlin, {\it ibid. } 22, 2695 (1980);
                                             31, 213(E) (1985);
                A. Sirlin and W. J. Marciano, Nucl. Phys. B189, 442 (1981);
                K. I. Aoki et al. , Prog. Theor. Phys. Suppl. 73, 1 (1982). 
\item[{\rm [12]}] H. L. Lai et. al. , Phys. Rev. D51, 4763 (1995). 
\item[{\rm [13]}] A. D. Martin, R. G. Roberts and W. J. Stirling,
		 Phys. Lett. B354, 155 (1995). 
\item[{\rm [14]}] G. Passarino and M. Veltman, Nucl. Phys. B160, 151(1979). 
\item[{\rm [15]}]  For a review see, for example, K. W. Merritt, 
                   hep-ex/9701009. 
\end{description}
\end{itemize}
\eject

\begin{center} {\Large Figure Captions} \end{center}
\vspace{.7cm}

Fig.1 Feynman diagrams for single top quark production via
$q \bar q'\rightarrow W^* \rightarrow t\bar b$:
(a) the tree-level diagram,  and (b)-(g) the one-loop self-energy 
and vertex SUSY QCD corrections.

Fig.2  The SUSY QCD correction $\Delta\sigma/\sigma_0$  
as a function of  gluino mass, assuming a squark mass of 100GeV.

Fig.3 The SUSY QCD correction $\Delta\sigma/\sigma_0$  
as a function of  squark mass, assuming a gluino mass of 200GeV.

\end{document}